%% file: main.tex
\documentclass[acmsmall]{acmart}

\setcopyright{rightsretained}
\acmJournal{TOSEM}
\acmYear{2023}
\acmVolume{1}\acmNumber{1}
\acmArticle{1}
\acmMonth{1}
\acmPrice{}
\acmDOI{10.1145/3624742}

\usepackage[normalem]{ulem}
\usepackage[utf8]{inputenc}
\usepackage{amsmath}  
\usepackage{graphicx}
\usepackage{multirow}
\usepackage{enumitem}
\usepackage{array}
\usepackage{tcolorbox}
\usepackage{xcolor}
\usepackage{balance}
\usepackage{mathtools}
\usepackage{xspace}
\usepackage{url}
\usepackage{color}
\usepackage{float}
\usepackage{xcolor}
\usepackage{amsmath}

\newcommand{\ptov}[0]{Post2Vec}
\newcommand{\so}{{Stack Overflow}\xspace}

\begin{document}
\title{Representation Learning for Stack Overflow Posts: How Far are We?}


\author{Junda He}
\email{jundahe@smu.edu.sg}
\affiliation{%
  \institution{Singapore Management University}
  \streetaddress{80 Stamford Rd.}
  \postcode{178902}
  \city{Singapore}
  \country{Singapore}
}

\author{Xin Zhou}
\email{xinzhou.2020@phdcs.smu.edu.sg}
\affiliation{%
  \institution{Singapore Management University}
  \streetaddress{80 Stamford Rd.}
  \postcode{178902}
  \city{Singapore}
  \country{Singapore}
}

\author{Bowen Xu}
\authornote{Corresponding Authors}
\email{bxu22@ncsu.edu}
\affiliation{%
  \institution{North Carolina State University}
  \city{Raleigh}
  \country{USA}
}

\author{Ting Zhang}
\email{tingzhang.2019@phdcs.smu.edu.sg}
\affiliation{%
  \institution{Singapore Management University}
  \streetaddress{80 Stamford Rd.}
  \postcode{178902}
  \city{Singapore}
  \country{Singapore}
}

\author{Kisub Kim}
\authornotemark[1]
\email{kisubkim@smu.edu.sg}
\affiliation{%
 \institution{Singapore Management University}
  \streetaddress{80 Stamford Rd.}
  \postcode{178902}
  \city{Singapore}
  \country{Singapore}
}

\author{Zhou Yang}
\email{zyang@smu.edu.sg}
\affiliation{%
 \institution{Singapore Management University}
  \streetaddress{80 Stamford Rd.}
  \postcode{178902}
  \city{Singapore}
  \country{Singapore}
}

\author{Ferdian Thung}
\email{ferdianthung@smu.edu.sg}
\affiliation{%
  \institution{Singapore Management University}
  \streetaddress{80 Stamford Rd.}
  \postcode{178902}
  \city{Singapore}
  \country{Singapore}
}

\author{Ivana Clairine Irsan}
\email{ivanairsan@smu.edu.sg}
\affiliation{%
 \institution{Singapore Management University}
  \streetaddress{80 Stamford Rd.}
  \postcode{178902}
  \city{Singapore}
  \country{Singapore}
}

\author{David Lo}
\email{davidlo@smu.edu.sg}
\affiliation{%
  \institution{Singapore Management University}
  \streetaddress{80 Stamford Rd.}
  \postcode{178902}
  \city{Singapore}
  \country{Singapore}
}

\renewcommand{\shortauthors}{He et al.}

\input{sections/abstract}


\maketitle
\input{sections/intro.tex}

\input{sections/models}
\input{sections/tasks.tex}
\input{sections/setting}
\input{sections/result}
\input{sections/discussion}
\input{sections/relate}
\input{sections/conclusion}

\section{Acknowledgement}
This research / project is supported by the National Research Foundation, Singapore, under its Industry Alignment Fund – Pre-positioning (IAF-PP) Funding Initiative. Any opinions, findings and conclusions or recommendations expressed in this material are those of the author(s) and do not reflect the views of National Research Foundation, Singapore.

\bibliographystyle{ACM-Reference-Format}
\bibliography{reference}
\balance

\end{document}

%% file: sections/abstract.tex
\begin{abstract}
The tremendous success of Stack Overflow has accumulated an extensive corpus of software engineering knowledge, thus motivating researchers to propose various solutions for analyzing its content.
The performance of such solutions hinges significantly on the selection of representation models for Stack Overflow posts.
As the volume of literature on Stack Overflow continues to burgeon, it highlights the need for a powerful Stack Overflow post representation model and drives researchers' interest in developing specialized representation models that can adeptly capture the intricacies of Stack Overflow posts. The state-of-the-art (SOTA) Stack Overflow post representation models are Post2Vec and BERTOverflow, which are built upon neural networks such as convolutional neural network (CNN) and transformer architecture (e.g., BERT). 
Despite their promising results, these representation methods have not been evaluated in the same experimental setting. To fill the research gap, we first empirically compare the performance of the representation models designed specifically for Stack Overflow posts (Post2Vec and BERTOverflow) in a wide range of related tasks, i.e., tag recommendation, relatedness prediction, and API recommendation.
The results show that Post2Vec cannot further improve the state-of-the-art techniques of the considered downstream tasks, and BERTOverflow shows surprisingly poor performance.
To find more suitable representation models for the posts, we further explore a diverse set of transformer-based models, including (1) general domain language models (RoBERTa, Longformer, GPT2) and (2) language models built with software engineering-related textual artifacts (CodeBERT,  GraphCodeBERT, seBERT, CodeT5, PLBart, and CodeGen).
This exploration shows that models like CodeBERT and RoBERTa are suitable for representing Stack Overflow posts.
However, it also illustrates the ``No Silver Bullet'' concept, as none of the models consistently wins against all the others.
Inspired by the findings, we propose SOBERT, which employs a simple yet effective strategy to improve the representation models of Stack Overflow posts by continuing the pre-training phase with the textual artifact from Stack Overflow. 
The overall experimental results demonstrate that SOBERT can consistently outperform the considered models and increase the state-of-the-art performance significantly for all the downstream tasks. 
\end{abstract}

%% file: sections/intro.tex
\section{Introduction}
\label{sec:intro}
Serving as the most popular software question and answer (SQA) forum, \so (SO) has dramatically influenced modern software development practice. As of August 2023, the forum has accumulated more than 23 million questions and 35 million answers\footnote{\url{https://stackexchange.com/sites?view=list\#traffic}}. \so is broadly recognized as an invaluable knowledge base and supplemental resource for the software engineering (SE) domain~\cite{ptm4tag,xu2018prediction,Wei2022CLEARCL, biker}, which triggered the increased interest of researchers and software developers in a wide range of \so post-related tasks, for example, recommendation of post tags (aka. tag recommendation)~\cite{ptm4tag}, recommendation of APIs according to a natural language query (aka. API recommendation)~\cite{biker}, and the identification of related posts (aka. relatedness prediction)~\cite{xu2018prediction}. 

An essential step in yielding promising results for these \so-related tasks is to obtain suitable representations of the posts. A beneficial \so representation model can capture the semantic concept of the posts and reveal more explanatory features from the hidden dimensions. As the volume of SE literature on solving \so-related tasks~\cite{ptm4tag,xu2018prediction, Wei2022CLEARCL} continues to burgeon, it has underscored the demand for a quality \so representation.

Over the years, numerous representation models have been specifically proposed for modeling \so posts.
Xu et al. proposed Post2Vec~\cite{post2vec}, a CNN-based~\cite{lecun2015deep} representation model that leverages the tags of a post to guide the learning process and models the post as the combination of three complementary components (i.e., title, description, and code snippet). Their experimental results demonstrate that it can substantially boost the performance for a wide range of \so posts-related tasks~\cite{Shirani2019QuestionRO, ahasanuzzaman2018classifying, biker}. Tabassum et al.~\cite{bertoverflow} leveraged the more advanced transformer architecture and pre-trained BERTOverflow based on 152 million sentences from \so. The results demonstrate that the embeddings generated by BERTOverflow have led to a significant improvement over other off-the-shelf models (e.g., ELMo~\cite{elmo} and BERT~\cite{bert}) in the software named entity recognition (NER) task.

Although these existing \so-specific methods have been proven to be beneficial, the effectiveness of Post2Vec is only evaluated
on limited solutions (i.e., Support Vector Machine~\cite{xu2018prediction} and Random Forest~\cite{beyer2018automatically}) and BERTOverflow only experimented for the NER task. These motivate us to further study the performance of existing \so-specific representation models on a diverse set of tasks. Unexpectedly, we found that both Post2Vec and BERTOverflow perform poorly. Such findings motivate us to explore the effectiveness of a larger array of representation techniques in modeling \so posts.

In addition to the aforementioned \so-specific representation models, we further consider nine transformer-based language models that could be potentially suitable for post representation learning. These models can be classified into two types: SE domain-specific models and general domain models.
SE domain-specific models are trained with SE-related contents (i.e., Github repositories) and are suitable for capturing the semantics of technical jargon of the SE domain. We consider six SE domain-specific models: CodeBERT~\cite{codebert}, GraphCodeBERT~\cite{GraphCodeBERT}, seBERT~\cite{sebert}, CodeGen~\cite{nijkamp2022codegen}, CodeT5~\cite{codeT5}, and PLBart~\cite{plbart}. 
We also include models from the general domain as they are usually trained with a more diverse amount of data than domain-specific models. For general domain models, we consider RoBERTa~\cite{roberta}, Longformer~\cite{longformer}, and GPT2~\cite{gpt2}.

We evaluate the performance of the aforementioned representation models on multiple  \so-related downstream tasks (i.e., tag recommendation~\cite{post2vec}, API recommendation~\cite{biker}, and relatedness prediction~\cite{xu2018prediction}). Furthermore, we build SOBERT, a stronger transformer-based language model for modeling \so posts. 
Our experimental results reveal several interesting findings:

\begin{enumerate}[leftmargin=*]
\item \textit{Existing \so post representation techniques fail to improve the SOTA performance of considered tasks.} Xu et al. demonstrated that the addition of the feature vectors generated by Post2Vec is beneficial for improving the post representation for traditional machine learning techniques. However, we discover that appending the feature vectors from Post2Vec~\cite{post2vec} does not derive a beneficial effect on considered deep neural networks. 
Furthermore, we reveal that the embedding generated by BERTOverflow could only achieve reasonable performance in the API recommendation task and give surprisingly poor performance in the tag recommendation task.

\item \textit{Among all the considered models, none of them can always perform the best.} According to our experiment results, although several representation models can outperform the SOTA approaches, none can always perform the best. 
As a result, this motivates us to propose a new model for representing \so posts.
\item \textit{Continued pre-training based on \so textual artifact develops a consistently better representation model.} We propose SOBERT by further pre-training with \so data. The overall results show that SOBERT consistently boosts the performance in all three considered tasks, implying a better representation.
\end{enumerate}  
Overall, we summarize the contributions of our empirical study as follows: 
\begin{enumerate}[leftmargin=*]
\item We comprehensively evaluate the effectiveness of eleven representation models for \so posts in three downstream tasks.
\item We propose SOBERT by pre-training based on posts from 
\so and show that SOBERT consistently outperforms other representation models in multiple downstream tasks.
\item We derive several insightful lessons from the experimental results to the software engineering community.
\end{enumerate}

The rest of the paper is organized as follows. Section \ref{sec:models} categorizes representation learning models into three groups and briefly describes them.
We formulate the downstream tasks (i.e., tag recommendation, API recommendation, relatedness prediction) and their corresponding state-of-the-art method in Section \ref{sec:tasks}.
Section \ref{sec:settings} introduces our research questions and the experiment settings. 
In Section \ref{sec:result}, we answer the research questions and report the experiment results. 
Section \ref{sec:discussion} further analyzes the result and elaborates the insights with evidence. 
Section \ref{sec:relate} describes related works, and Section \ref{sec:conclusion} summarizes this study.

%% file: sections/models.tex
\section{Representation Learning Models}
\label{sec:models}
In this section, we summarize the considered representation models in this paper. We explore a wide range of techniques across the spectrum of representing \so posts, including two \so-specific post representation models (Post2Vec~\cite{post2vec} and BERTOverflow~\cite{bertoverflow}), six SE domain-specific transformer-based pre-trained representation models (PTM) (CodeBERT~\cite{codebert}, GraphCodeBERT~\cite{GraphCodeBERT}, seBERT~\cite{sebert}, CodeT5~\cite{codeT5}, PLBart~\cite{plbart}, and CodeGen~\cite{nijkamp2022codegen}) and three transformer-based PTMs from the general domain (RoBERTa~\cite{roberta} Longformer~\cite{longformer}, and GPT2~\cite{gpt2}).

\subsection{Transformer-based Language Models}

Transformer-based language models have revolutionized the landscape of representation learning in natural language processing (NLP)~\cite{bert, roberta, gpt2}. Their efficacy in capturing text semantics has led to unparalleled performance in various applications, such as sentiment analysis~\cite{bert_sentiment}, POS tagging~\cite{bert_pos}, and question answering~\cite{bert_questionanswering}. The vanilla transformer architecture~\cite{transformer} is composed of the encoder and decoder components. Based on the usage of these components, transformer-based language models can be categorized into three types: encoder-only, decoder-only, and encoder-decoder models.  

\textit{Encoder-only models} exclusively leverage the encoder stacks of the vanilla transformer~\cite{transformer} architecture. 
BERT~\cite{bert} stands as a prominent encoder-only representation model, which learns a bidirectional contextual representation of text. BERT proposes the \textit{Masked Language Modeling} (MLM) task at the pre-training phase. In MLM, the input data is corrupted by randomly masking 15\% of the tokens, and then the BERT model learns to reconstruct the original data by predicting the masked words. BERT is extensively pre-trained on large-scale datasets, which learn a meaningful representation that is reusable for various tasks, thus eliminating the process of training language models from scratch and saving time and resources. 

In contrast, \textit{Decoder-only models} consist solely of the decoder components of the original transformer architecture. A notable instance of such models is the GPT~\cite{gpt},
GPT operates under a causal language modeling (CLM) framework during its training phase. CLM is a strategy where the model predicts the next token in a sequence while only considering preceding tokens. In other words, this design restricts the model from accessing future tokens in the sequence.

Bridging the above approaches, textit{Encoder-decoder models} integrate both the encoder and decoder components of the transformer architecture. Popular encoder-decoder models involve T5~\cite{raffel2020exploring} and BART~\cite{bart}. 
The T5 model~\cite{raffel2020exploring} advocates a unified text-to-text framework that converts various language tasks into a consistent text-to-text format. T5 is pre-trained on the Colossal Clean Crawled Corpus~\cite{raffel2020exploring}, along with a mixture of unsupervised and supervised pre-training tasks. On the other hand, BART~\cite{bart} introduces a variety of noising functions to corrupt the initial input sequence (i.e., token deletion, document rotation, and sentence shuffling) during the pre-training phase. By corrupting the original sequence through these mechanisms, BART is trained to restore the original input.

\subsection{Existing Representation Models for \so Posts}

\noindent\textbf{\ptov{}}~\cite{post2vec}
is the latest approach proposed specifically for \so post representation learning~\cite{post2vec}.
Post2Vec is designed with a \emph{triplet} architecture to process three components of a \so post (i.e., title, text, and code snippets). It leverages Convolutional Neural Networks (CNNs) as feature extractors to encode the three components separately. 
The corresponding three output feature vectors are then fed to a feature fusion layer to represent the post. 
In the end, \ptov{} uses tag information of the post, which is considered as the post's general semantic meaning to supervise the representation learning process. 
Xu et al. demonstrated that the representation learned by \ptov{} can enhance the feature vectors for \so-related downstream tasks (e.g., relatedness prediction and API recommendation). 
For each downstream task, the vector representation learned by \ptov{} is combined with the feature vector produced by the corresponding state-of-the-art approach to form a new feature vector. The new feature vector is used to boost the performance of the corresponding model for the task.
Following the experiment settings of Xu et al., we use Post2Vec as a complementary feature vector to the state-of-the-art approach in this paper. Specifically, we concatenate the post representation generated by Post2Vec to the original feature vector of the state-of-the-art approach. This combined feature vector is employed in further training.  

\noindent\textbf{BERTOverflow}~\cite{bertoverflow} keeps the original BERT architecture, and it leverages 152 million sentences and 2.3 billion tokens from \so to pre-train \so-specific word embeddings. The authors have leveraged the embedding generated by BERTOverflow to implement a software-related named entity recognizer (SoftNER). The performance of SoftNER is experimented with the name entity recognition (NER) task for the software engineering domain, focusing on identifying code tokens or programming-related named entities that appear within SQA sites like \so. The results show that BERTOverflow outperforms all other models in the proposed task.

\subsection{Representation Models from SE domain}

\noindent\textbf{CodeBERT}~\cite{codebert}
is a SE knowledge-enriched bi-modal pre-trained model, which is capable of modeling both natural languages (NL) and programming languages (PL).The CodeBERT model has shown great effectiveness in a diverse range of SE domain-specific activities, for example, code search~\cite{codebert}, traceability prediction~\cite{tracebert}, and code translation~\cite{codetranslate}. CodeBERT inherits the architecture of BERT~\cite{bert}, and it continues pre-training based on the checkpoint of RoBERTa~\cite{roberta} with the NL-PL data pairs obtained from the CodeSearchNet dataset~\cite{codesearchnet}.
It has two popular pre-training objectives: Masked Language Modeling (MLM) and Replaced Token Detection (RTD)~\cite{electra}. Rather than masking the input like MLM, RTD corrupts the input by replacing certain tokens with plausible substitutes. RTD then predicts if each token in the altered input was replaced or remained unchanged.
The eventual loss function for CodeBERT at the pre-training stage is the combination of both MLM and RTD objectives, where $\theta$ denotes the model parameters:
\begin{equation}
    \underset{\theta}{\text{mini}}( \mathcal{L}_{RTD}(\theta)+\mathcal{L}_{MLM}(\theta))
\end{equation}

\noindent\textbf{GraphCodeBERT}~\cite{GraphCodeBERT}
incorporates a hybrid representation in source code modeling. Apart from addressing the pre-training process over NL and PL, GraphCodeBERT utilizes the data flow graph of source code as additional inputs and proposes two structure-aware pre-training tasks (i.e., Edge Prediction and Node Alignment) aside from the MLM prediction task. GraphCodeBERT is evaluated in code search~\cite{codebert}, clone detection~\cite{codeclone}, code translation~\cite{codetranslate}, and code refinement~\cite{coderefine}, respectively. It outperforms CodeBERT and all the other baselines, including RoBERTa (code version)~\cite{GraphCodeBERT}, Transformer~\cite{transformer}, and LSTM~\cite{lstm}.

\noindent\textbf{seBERT}~\cite{sebert} aims to advance the previous PTMs in the SE context with a larger model architecture and more diverse pre-training data. The authors pre-trained seBERT using the BERT$_{LARGE}$ architecture, i.e., with 24 layers, a hidden layer size of 1024, and 16 self-attention heads, with a total of 340 million parameters. seBERT is pre-trained with more than 119GB of data from four data sources, i.e., \so posts, GitHub issues, Jira issues, and Github commit messages. The model's effectiveness is verified in three classification tasks, i.e., issue type prediction, commit intent prediction, and sentiment mining. The authors observe the experiment results showing that seBERT is significantly better than BERToverflow in these tasks.

\noindent\textbf{CodeGen}~\cite{nijkamp2022codegen} is a pre-eminent decoder-only transformer-based PTM for program synthesis, it undergoes pre-training on an extensive dataset comprising both natural language and programming languages. This model innovates a multi-turn programming synthesis paradigm and creates a comprehensive benchmark for multi-turn programming tasks.
The multi-turn program synthesis approach involves users and the model together in multiple steps.
The user communicates with the model by progressively providing specifications in natural language
while receiving responses from the model in the form of synthesized subprograms. 
CodeGen demonstrates state-of-the-art performance on Python code generation on HumanEval~\cite{humaneval} and a set of tasks in the multi-turn programming benchmark. The model is pre-trained on BigQuery dataset\footnote{\url{https://cloud.google.com/bigquery/public-data}}. This dataset contains GitHub repositories with multiple programming languages, including C, C++, Go, Java, JavaScript, and Python.

\noindent\textbf{CodeT5}~\cite{codeT5} is an encoder-decoder PTM that is designed to better consider the code semantics conveyed from the identifiers from code. It has the same architecture as T5~\cite{raffel2020exploring} and employs a multi-task training process to support both code understanding and generation tasks. CodeT5 utilizes a novel identifier-aware pre-training task that enables the model to distinguish which code tokens are identifiers and to recover them when they are masked. To improve the NL-PL alignment, CodeT5 further incorporates a bimodal dual learning objective
for a bidirectional conversion between natural languages and programming languages.

\noindent\textbf{PLBart}~\cite{plbart} is also an encoder-decoder PTM, which is capable of performing a broad spectrum of program-related understanding and generation tasks. 
PLBART employs the same architecture as BART~\cite{bart} and is pre-trained on a large collection of Java and Python functions and associated natural language documentation from Github repositories and \so posts. Experiments showed that PLBART can achieve promising performance in code summarization, code generation, and code translation.

\subsection{Representation Models from General Domain}

\noindent\textbf{RoBERTa}~\cite{roberta}
is a replication study on the pre-training objectives of BERT~\cite{bert}, and analyzed the impact of key hyper-parameters. The insights from the replication study have led to the development of RoBERTa, which is an improved version of BERT. In comparison with BERT, RoBERTa has made several modifications to the pre-training stage: (1) training with larger batch size, more data, and longer training time; (2) abandoning the next sentence prediction (NSP) task of BERT and showed that removal of NSP slightly improves the model efficiency; (3) training with longer sequences; (4) masking the training data dynamically rather than statically. 

\noindent\textbf{Longformer}~\cite{longformer}
aims to alleviate the limitation of transformer-based models in processing long sequences.  
The self-attention mechanism of the transformer suffers from the $O(n^2)$ quadratic computational complexity problem, which restricts the ability of transformer-based models to model long sequences.
Pre-trained models like BERT~\cite{bert} and RoBERTa~\cite{roberta} only accept a maximum input of 512 tokens.
Longformer leverages a combination of sliding window attention and global attention mechanism such that the computational memory consumption scales linearly as the sequence becomes longer. In contrast to models like RoBERTa and CodeBERT, which could only accept a maximum of 512 tokens as input, Longformer supports sequences of length up to 4,096. Similar to CNN~\cite{lecun2015deep}, Longformer lets each input token only attend to surrounding neighbors that are within a fixed window size. Denoting the window size as $w$, each token could only attend to $ \frac{1}{2}w$ tokens on both sides, thus decreasing the computation complexity to $O(n \times w)$. 
However, the sliding window may compromise the performance as it cannot capture the whole context. To compensate for the side-effect, global tokens are selected. Such tokens are implemented with global attention, which attends to all other tokens, and other tokens also attend to the global tokens.

\noindent\textbf{GPT2}~\cite{gpt2} is a decoder-only model. It is trained with a simple objective: predicting the next word, given all of the previous words within the text. GPT-2 is trained on a dataset of 8 million web pages. The diversity of the dataset causes this simple goal to contain naturally occurring demonstrations of many tasks across diverse domains. GPT-2 is a direct scale-up of GPT, with more parameters and trained on more than 10 times the amount of data. 

%% file: sections/tasks.tex
\section{Downstream Tasks}
\label{sec:tasks}
 In this section, we formulate the target problems that are used to measure the effectiveness of the representation models and then describe the corresponding state-of-the-art solution. We select multiple \so-related downstream tasks, which have been popular research topics for \so posts. 
To be more specific, we consider: \textit{Tag Recommendation}~\cite{ptm4tag, post2vec}, \textit{API Recommendation}~\cite{Wei2022CLEARCL,biker} and \textit{Relatedness Prediction}~\cite{xu2018prediction,Pei2021AttentionbasedMF}, covering a multi-label classification problem, a ranking problem, and multi-class classification problem. All selected tasks operate on the abstraction of a post, which could be benefited from a high-quality \so post representation.

\subsection{Tag Recommendation}
The user-annotated tags of a \so post serve as helpful metadata and have a critical role in organizing the contents of \so posts across different topics. Suitable tags precisely summarize the message of a post, while redundant tags and synonym tags make it more difficult in maintaining the content of the site.
A tag recommendation system could effectively simplify the tagging process and minimize the effect of manual errors, therefore, avoiding problems like tag synonyms and tag redundancy.
\subsubsection{Task Formulation}
We formulate the tag recommendation task as a \textit{multi-label classification problem}. Given $\mathcal{X}$ as the corpus of \so posts, and $\mathcal{Y}$ denotes the total collection of tags, we represent each post as $x_i$, where $0 \leq i \le |X|, i \in \mathbb{N}$ and the tags of each post as $y_i \subset \mathcal{Y}$. The goal is to recommend the most relevant set of tags $y_i$ to $x_i$. 

\subsubsection{State-of-the-art technique} PTM4Tag~\cite{ptm4tag} leverages three pre-trained models to solve the tag recommendation problem. The three pre-trained models are responsible for modeling the title, description, and code snippet, independently.

\subsection{API Recommendation}
Modern software development process heavily relies on third-party APIs, which leads to the research of an automated API recommendation approach that is intended to simplify the process of API search~\cite{Xia2017WhatDD}. 
Questions related to APIs are one of the most viewed topics on \so~\cite{Huang2018APIMR}. \so consists of an enormous amount of discussion about API usage. Developers are more intended to search for relevant \so posts and pick out the APIs that seem useful in the discussions~\cite{huang2018api} rather than checking API documentation. Thus, it makes \so the primary source for building a dataset of the API recommendation task.

\subsubsection{Task Formulation}
We follow the exact task definition as the previous literature~\cite{Wei2022CLEARCL,Huang2018APIMR,Gu2016DeepAL}. 
Given a natural language (NL) query that describes programming requirements, the goal is to recommend relevant APIs that implement the function for the query.
Thus, the task aims to inform developers which API to use for a programming task.
Formally speaking, given the corpus of natural language query $\mathcal{Q}$, we denote each query as $q_i$. The goal of the API recommendation system is to find a set of relevant APIs $y_i \subset \mathcal{Y}$ for $q_i$, where $\mathcal{Y}$ is the total set of available APIs.

\subsubsection{State-of-the-art technique} Wei et al.~\cite{Wei2022CLEARCL} proposed CLEAR, an automated approach that recommends API by embedding queries and \so posts with a PTM (distilled version of the RoBERTa\footnote{\url{https://huggingface.co/distilroberta-base}}). Given a NL query, CLEAR firstly picks a sub-set of candidate \so posts based on the embedding similarity to reduce the search space. Then, CLEAR ranks the candidate \so posts and recommends the APIs from the top-ranked \so posts. 

\subsection{Relatedness Prediction}
The notion of a knowledge unit (KU) is defined as a set containing a question along with all its answers~\cite{xu2018prediction, Pei2021AttentionbasedMF, Shirani2019QuestionRO}. To find a comprehensive technical solution for a given problem, developers usually need to summarize the information from multiple related KUs. However, searching for related KUs can be time-consuming as the same question can be rephrased in many different ways. Thus, researchers have proposed several techniques to automate the process of identifying the related KUs~\cite{xu2018prediction,Pei2021AttentionbasedMF,Shirani2019QuestionRO}, which could significantly improve the efficiency of the software development cycle.
\subsubsection{Task Formulation} 
 The task is commonly formulated as a multi-class classification problem~\cite{xu2018prediction,Pei2021AttentionbasedMF,Shirani2019QuestionRO}. The relatedness between questions is classified into four classes, from the most relevant to irrelevant, which are:
\begin{itemize}
    \item \textit{Duplicate}: Two KUs correspond to a pair of semantically equivalent questions. The answer of one KU can also be used to answer another KU. 
    \item \textit{Direct}: One KU is beneficial in answering the question in another KU, for example, by explaining certain concepts and giving examples.
    \item \textit{Indirect}: One KU provides relevant information but does not directly answer the questions of another KU.
    \item \textit{Isolated}: The two KUs are semantically uncorrelated.
\end{itemize}

Given the set \(K\) of all knowledge units, the goal of relatedness prediction is to predict the degree of relatedness between any two KUs, \(k_i\) and \(k_j\). The relatedness class is denoted as $C$, where $ C=\{\textit{Duplicate}, \textit{Direct}, \textit{Indirect}, \textit{Isolated} \}$. 
Formally, the task of relatedness prediction is defined as obtaining the function $R$, such that
$ R(k_i, k_j) = c$, where \( c \in C \).

In Table \ref{tab:relate-examples}, we demonstrate examples on pairs of KUs with different relatedness.
\begin{itemize}
    \item Original KU: This KU addresses the topic of string comparison in Java.
    \item Duplicate KU: This KU introduces the difference of "equals" and "==". Essentially, it offers a varied perspective on the same topic of string comparison in Java.
    \item Direct KU: This KU focuses on the behavior of "==" during string concatenation in Java. As it provides insights directly beneficial to understanding the original KU's topic without being a duplicate, it is categorized as directly related.
    \item Indirect KU: This KU discusses the memory allocation during string concatenation in Java. While it doesn't directly address the main topic of string comparison, its relevance to the direct KU concerning string operations classifies it as an indirectly related unit.
\end{itemize}

\begin{table}[]
    \centering
    \small
    \caption{Examples of duplicate, direct, indirect knowledge units pairs for the relatedness prediction task. }
    \label{tab:relate-examples}
    \begin{tabular}{c|c|c}
        \hline
                 & \textbf{Post ID}  & \textbf{Title}   \\ \hline
    \textbf{Original KU}  & 513832   & \textit{How do I compare strings in Java?}                                                                                                 \\
    \textbf{Duplicate KU} & 3281448  & \textit{Strings in Java : equals vs ==}                                                                                                    \\
    \textbf{Direct KU}    & 34509566 & \textit{"==" in case of String concatenation in Java}                                                                                     \\
    \textbf{Indirect KU}  & 11989261 & \begin{tabular}[c]{@{}c@{}}
    \textit{Does concatenating strings in Java always} \\ 
    \textit{lead to new strings being created in memory?}
    \end{tabular}
    \\
    \hline
    \end{tabular}
    \end{table}
\subsubsection{State-of-the-art technique} Recently, Pei et al. introduced ASIM~\cite{Pei2021AttentionbasedMF}, which yielded state-of-the-art performance in the relatedness prediction task. Pei et al. pre-trained word embeddings specialized to model \so posts with a corpus collected from the \so data dump. Then ASIM uses BiLSTM~\cite{schuster1997bidirectional} to extract features from \so posts and implements the attention mechanism to capture the semantic interaction among the KUs. 

%% file: sections/setting.tex
\section{Research Questions and Experimental Settings}
\label{sec:settings}
In this section, we first introduce our research questions and then describe the corresponding experiment settings.

\subsection{Research Questions}

\subsubsection*{\textbf{RQ1. How effective are the existing \so post representation models?}}

Various methods have been proposed in modeling \so posts. However, there is still a lack of analysis of the existing \so-specific representation methods. For instance, Xu et al.~\cite{post2vec} have demonstrated that Post2Vec is effective in boosting the performance of traditional machine learning algorithms, i.e., support vector machine (SVM) and Random Forest. However, the efficacy of Post2Vec in facilitating deep learning-based models has not yet been investigated. Moreover, Tabassum et al.~\cite{bertoverflow} only leveraged the embeddings from BERTOverflow in the software-related NER task, but not for other popular \so-related tasks. In light of this research gap, we aim to evaluate the current \so-specific representation methods for popular \so-related tasks under the same setting for this research question. 

\subsubsection*{\textbf{RQ2. How effective are the popular transformer-based language models for the targeted downstream tasks?}}

In addition to the existing \so representation models, we explore the effectiveness of a wider spectrum of representation models. 
transformer-based language models have shown great performance and generalizability in representation learning. 
Representations generated by such models have demonstrated promising performance in a broad range of tasks with datasets of varying sizes and origins.
Borrowing the best-performing representation models from various domains and investigating their performance can derive interesting results, as recent literature~\cite{yang2022aspect,zhang2020sentiment} have revealed that they are potentially great candidates for representing posts as well. 
This motivates us to employ RoBERTa~\cite{roberta} and Longformer~\cite{longformer} from the general domain and CodeBERT~\cite{codebert}, GraphCodeBERT~\cite{GraphCodeBERT}, and seBERT~\cite{sebert} from the SE domain.
We set up the exact same experimental settings for each model.

\subsubsection*{\textbf{RQ3. Is further pre-training on \so data helpful in building a better model?}}

Further pre-trained models with domain-specific corpus have been common practice in the NLP domain, however, their effectiveness is not verified for representing \so posts. 
In this RQ, we introduce SOBERT, which is obtained by continuing the pre-training process on CodeBERT with \so data, and we aim to investigate whether further pre-training with \so data improves the performance. 

\subsection{Experimental Settings}
\subsubsection{\textbf{Tag Recommendation}}
\label{setting:tag}
\subsubsection*{\textbf{Dataset}}
\hfil \\
The dataset used by He et al.~\cite{ptm4tag} in the training of PTM4Tag only includes the \so posts dated before September 5, 2018. To address this limitation, we use the \so data dump released in August of 2022 to construct a new dataset for our experiment. Ideally, a tag recommendation approach should only learn from high-quality questions. Therefore, we remove the low-quality questions when constructing the dataset.
According to the classification criteria of question quality defined by Ponzanelli et al.~\cite{PonzanelliMBLF14}, we first filter out the questions that do not have an accepted answer and further remove the questions with a score of less than 10.
Additionally, we exclude the rare tags and rare posts. Previous literature in tag recommendation~\cite{ptm4tag, post2vec} has defined a tag as rare if it occurs less than 50 times within the dataset, and a post is considered rare if all of its tags are rare tags. The usage of rare tags is discouraged since it implies the unawareness of the tag among developers. We follow the same definition as the previous literature and set the frequency threshold for rare tags as 50. 
In the end, the resultant dataset consists of 527,717 posts and 3,207 tags. We split the dataset into a training set, a validation set, and a test set according to the 8:1:1 ratio, which corresponds to 422,173, 52,772, and 52,772 posts, respectively.

During the training process, we only consider the question posts from \so and ignore the answer posts. We check the post-IDs of the question posts in our dataset to ensure that each post has a unique post-ID. 
Each question post consists of two components, which are the title and body. The code snippets within the body of a post are enclosed in HTML tags \texttt{<pre><code>} and \texttt{</code></pre>}, we cleaned the redundant HTML tags with regular expression, we first clean the HTML tag by using regular expressions \texttt{<pre><code>([\textbackslash{}s\textbackslash{}S]*?)<//code><//pre>}.
After that, we concatenate the title and body together to form the final input data. Thus, our input data has integrated the title, natural languages in the body, and code snippets in the body. This input data is then fed into the tokenizer of the representation model to be tokenized.

\subsubsection*{\textbf{Evaluation Metrics}}
\hfil \\
We report the performance for this task using Precision@k, Recall@k, and F1-score@k, where k indicates the top-k recommendations. Such metrics are extensively used in previous works~\cite{ptm4tag, post2vec, tagdc, zhou2019deep}, and we calculate the average score for each of them. Mathematically speaking, the evaluation metrics are computed as follows:
$$
 Precision@k =
\frac{  \lvert  \text{Tag}_{\text{True} } \cap \text{Tag}_{ \text{Predict} } \rvert }{k}  
$$

$$
Recall@k_i =   \begin{cases}
      \frac{| \text{Tag}_{\text{True}} \cap \text{Tag}_{\text{Predict}}}  {k}| & \text{if } | \text{Tag}_{\text{True}}| > 
      
      k\\
      \frac{| \text{Tag}_{\text{True}} \cap \text{Tag}_{\text{Predict}} | }{|\text{Tag}_{\text{True}}|} & \text{if } |\text{Tag}_{\text{True}}| \leq k\\
    \end{cases} \\
$$
$$
F1\text{-}score@k =  2 \times \frac{ Precision@k  \times Recall@k }{ Precision@k  + Recall@k }$$

\noindent In the above formulas, $\text{Tag}_{\text{True}}$ refers to the ground truth tags and $\text{Tag}_{\text{Predict}}$ refers to the predicted tags. 
Notice that the above formula of Recall@k is determined by conditions since Recall@k naturally disfavors small k. The revisited Recall@k has been widely adopted in previous experiments of tag recommendation~\cite{ptm4tag, post2vec,zhou2019deep}. Since \so posts cannot have more than 5 tags, we report the results by setting the k as 1, 3, and 5.
\subsubsection*{\textbf{Implementation Details}}\hfil \\

For Longformer, we set the maximum accepted input sequence as 1,024, and for other transformer-based language models the maximum input sequence is set as 512. This setting of the input sequence is kept the same for the other two tasks (API recommendation and relatedness prediction). For encoder-decoder and decoder-only representation models, we select the last
decoder hidden state as the representation following previous literature~\cite{codeT5, bart}. 

We set the learning rate as 5e-5, batch size as 512, epoch number as 30, and use the Adam optimizer to update the parameters. 
We save the model at the end of each epoch and select the model with the smallest validation loss to run the evaluation.  


\subsubsection{\textbf{API Recommendation}}
\subsubsection*{\textbf{Dataset}}\hfil \\
We use the BIKER dataset~\cite{huang2018api} crafted by Huang et al., which is the same dataset used by Wei et al.~\cite{Wei2022CLEARCL}. The training set contains 33K questions with corresponding relevant APIs in the accepted answers. 
The test dataset contains manually labeled questions from \so, which are looking for API to solve programming problems and labeled the ground-truth API for these questions based on their accepted answers. 

When creating the Biker dataset, Huang et al. first select question posts from \so satisfying the following three criteria: (1) the question has a positive score, (2) at least one answer to the question contains API entities (3) the answer has a positive score. Huang et al. then manually inspected the collected questions and removed the questions that were not about searching APIs for programming tasks.
 
Huang et al. aim to create queries that do not have too many words, thus only the titles of the \so posts are used as queries. The ground truth APIs were extracted from the code snippets in the accepted answers in filtered posts. Again, the extracted APIs were manually checked to ensure their correctness. Eventually, the test dataset contains 413 questions along with their ground truth APIs after the manual labeling process. 
The titles of these questions are used as the queries for API searching.


\subsubsection*{\textbf{Evaluation Metrics}}\hfil \\
We use the same evaluation metrics as previous literature~\cite{Huang2018APIMR,Wei2022CLEARCL} for the API recommendation task. The metrics are Mean Reciprocal Rank (MRR), Mean Average Precision (MAP), Precision@k and Recall@k. Mathematically, MRR is defined as:
$$MRR = \frac{1}{|Q|} \sum_{i=1}^{|Q|} \frac{1}{rank_i} $$
where $Q$ refers to all queries and $rank_i$ refers to the rank position of the first relevant API from the recommended API list for the query$_i$. For a given query, MRR gives a score of $\frac{1}{rank}$, where $rank$ is the ranking of the first correct API in the recommended list. In other words, the score of MRR is inversely proportional to the rank of the first correct API. MAP is the mean of average precision scores ($AveP$) for each query. While MRR gives the score based on the ranking of the first correct answer, MAP considers the ranks of all correct answers to measure the quality of the recommended list. 
Mathematically, MAP is defined as:
$$MAP = \frac{1}{|Q|} \sum_{i=1}^{|Q|} AveP(i) $$
$AveP(i)$ itself is defined as:
$$AveP (i) = \frac{1}{|K|} \sum_{k \in K} \frac{num(k)}{k}$$
where $K$ is the set of ranking positions of the relevant
APIs from the APIs list of query$_i$, and
$num(k)$ represents the number of relevant API in the top-k.

Differently from tag recommendation, the Recall@k metrics used in this task follow the conventional definition, which is:
$$ Recall@k = \frac{|\text{API}_{\text{True}} \cap \text{API}_{\text{Predict}}|}{|\text{API}_{\text{True}}|} $$
To be consistent with Wei et al.~\cite{Wei2022CLEARCL}, we use k $\in {1,3,5}$.

\subsubsection*{\textbf{Implementation Details}}\hfil \\
CLEAR shows state-of-the-art performance in the API recommendation task by leveraging BERT sentence embedding and contrastive learning. The original architecture of CLEAR is implemented based on DistilRoBERTa \footnote{\url{https://huggingface.co/distilroberta-base}} during the training process.
In this study, we also explore the effectiveness of other representation methods by replacing the embedding of DistilRoBERTa in CLEAR. 
For Post2Vec, we concatenate the post representation from Post2Vec to the original implementation of CLEAR.

For this task, we set the batch size as 256, and the epoch number as 30. Same to the description in Sec \ref{setting:tag}, we select the model with the smallest validation loss to run the test set. 

\subsubsection{\textbf{Relatedness Prediction}}
\subsubsection*{\textbf{Dataset}}\hfil \\
The experiments are conducted based on the KUs dataset provided by Shirani et al.~\cite{Shirani2019QuestionRO}. This dataset\footnote{\url{https://anonymousaaai2019.github.io}} contains 34,737 pairs of KUs. To ensure a fair comparison with the prior work~\cite{Pei2021AttentionbasedMF}, we use the same data for training, validation, and testing, containing 208,423, 34,737, and 104,211 pairs of KU, respectively.
Our input data is the concatenation of two KUs. Specifically, 
each KU contains one question post and three corresponding answer posts. For the question post, we include the title, body, and code snippets. For the answer post, we consider the body and code.

\subsubsection*{\textbf{Evaluation Metrics}}\hfil \\
Following prior work~\cite{Pei2021AttentionbasedMF},
we adopt the micro-averaging method to calculate Micro-precision, Micro-recall, and Micro-F1 as evaluation metrics.

\subsubsection*{\textbf{Implementation Details}}\hfil \\
We concatenate a pair of posts as the input to train a multi-class classifier. We fine-tuned Longformer on a sequence length of 1,024 and fine-tuned other pre-trained models on a sequence length of 512. 
For all experiments, we set the batch size as 32 and the epoch number as 5. We select the model with the smallest validation loss to run the evaluation.  

%% file: sections/result.tex
\section{Experimental Results}
\label{sec:result}

This section describes the experiment results and answers our research questions. The experimental results are summarized in Table \ref{tab:tag}, \ref{tab:api}, and \ref{tab:relate}, respectively.

\begin{table}
    \centering
    \caption{Experiment Results for Tag Recommendation Task}
    \label{tab:tag}
    \resizebox{\linewidth}{!}{%
    \begin{tabular}{c|c|c|c|c|c|c|c|c|c|c} 
    \hline
    \textbf{Group}                                                                               & \textbf{Representation} & \textbf{P@1}   & \textbf{R@1}            & \textbf{F1@1}           & \textbf{P@3}   & \textbf{R@3}   & \textbf{F1@3}  & \textbf{P@5}   & \textbf{R@5}   & \textbf{F1@5}           \\ 
    \hline
    \textbf{SOTA}                                                                                & \textbf{PTM4Tag}        & 0.875          & 0.875                   & 0.875                   & 0.586          & 0.756          & 0.641          & 0.417          & 0.805          & 0.526                   \\ 
    \hline
    \multirow{2}{*}{\textbf{SO-Specific}}                                                        & \textbf{Post2Vec}       & 0.875          & 0.875                   & 0.875                   & 0.585          & 0.754          & 0.639          & 0.416          & 0.804          & 0.525                   \\ 
    \cline{2-11}
                                                                                                 & \textbf{BERTOverflow}   & 0.088          & 0.088                   & 0.088                   & 0.089          & 0.094          & 0.095          & 0.083          & 0.163          & 0.105                   \\ 
    \hline
    \multirow{3}{*}{\begin{tabular}[c]{@{}c@{}}\textbf{General }\\\textbf{ Domain}\end{tabular}} & \textbf{RoBERTa}        & 0.878          & 0.878                   & 0.878                   & 0.591          & 0.761          & 0.646          & 0.418          & 0.804          & 0.527                   \\ 
    \cline{2-11}
                                                                                                 & \textbf{Longformer}     & 0.852          & 0.852                   & 0.852                   & 0.559          & 0.721          & 0.612          & 0.397          & 0.769          & 0.502                   \\ 
    \cline{2-11}
                                                                                                 & \textbf{GPT2}           & 0.884          & 0.884                   & 0.884                   & 0.593          & 0.763          & 0.648          & 0.418          & 0.805          & 0.528                   \\ 
    \hline
    \multirow{6}{*}{\begin{tabular}[c]{@{}c@{}}\textbf{SE }\\\textbf{ Domain}\end{tabular}}      & \textbf{CodeBERT}       & 0.876          & 0.876                   & 0.876                   & 0.588          & 0.758          & 0.642          & 0.418          & 0.805          & 0.527                   \\ 
    \cline{2-11}
                                                                                                 & \textbf{GraphCodeBERT}  & 0.874          & 0.875                   & 0.875                   & 0.582          & 0.751          & 0.636          & 0.410          & 0.791          & 0.517                   \\ 
    \cline{2-11}
                                                                                                 & \textbf{seBERT}         & 0.088          & 0.088                   & 0.088                   & 0.089          & 0.094          & 0.095          & 0.083          & 0.163          & 0.105                   \\ 
    \cline{2-11}
                                                                                                 & \textbf{CodeT5}         & 0.887          & 0.887                   & 0.887                   & 0.599          & 0.770          & 0.653          & 0.420          & 0.809          & 0.530                   \\ 
    \cline{2-11}
                                                                                                 & \textbf{PLBart}         & 0.883          & 0.883                   & 0.883                   & 0.600          & 0.773          & 0.656          & 0.422          & 0.811          & 0.532                   \\ 
    \cline{2-11}
                                                                                                 & \textbf{CodeGen}        & 0.872          & 0.872                   & 0.872                   & 0.584          & 0.751          & 0.638          & 0.411          & 0.792          & 0.519                   \\ 
    \hline
    \textbf{Our Model}                                                                           & \textbf{SOBERT}     & \textbf{0.905} & \textbf{\textbf{0.905}} & \textbf{\textbf{0.905}} & \textbf{0.615} & \textbf{0.790} & \textbf{0.671} & \textbf{0.437(+3.4\%)} & \textbf{0.836(+3.0\%)} & \textbf{0.551(+3.4\%)}  \\
    \hline
    \end{tabular}
    }
    \end{table}
    
    \begin{table*}
        \centering
        \caption{Experimental Results for API Recommendation Task}
        \label{tab:api}
        \resizebox{\linewidth}{!}{%
        \begin{tabular}{c|c|c|c|c|c|c|c|c|c} 
        \hline
        \textbf{Group}                           & \textbf{Representation} & \textbf{MRR}   & \textbf{MAP}   & \textbf{P@1}    & \textbf{P@3}   & \textbf{P@5}   & \textbf{R@1}   & \textbf{R@3}   & \textbf{R@5}                        \\ 
        \hline
        \textbf{SOTA}                            & \textbf{CLEAR}          & 0.739          & 0.753          & 0.482           & 0.560          & 0.562          & 0.629          & 0.766          & 0.793                               \\ 
        \hline
        \multirow{2}{*}{\textbf{SO-Specific}}    & \textbf{Post2Vec}       & 0.735          & 0.745          & 0.471           & 0.560          & 0.556          & 0.625          & 0.774          & 0.801                               \\ 
        \cline{2-10}
                                                 & \textbf{BERTOverflow}   & 0.753          & 0.778          & 0.521           & 0.639          & 0.651          & 0.681          & 0.774          & 0.762                               \\ 
        \hline
        \multirow{3}{*}{\textbf{General domain}} & \textbf{RoBERTa}        & 0.777          & 0.790          & 0.537           & 0.640          & 0.653          & 0.689          & 0.782          & 0.815                               \\ 
        \cline{2-10}
                                                 & \textbf{Longformer}     & 0.767          & 0.782          & 0.525           & 0.623          & 0.646          & 0.683          & 0.772          & 0.793                               \\ 
        \cline{2-10}
                                                 & \textbf{GPT2}           & 0.766 & 0.782 & 0.528 & 0.641 & 0.650 & 0.683 & 0.772 & 0.795  \\ 
        \hline
        \multirow{6}{*}{\textbf{SE domain}}      & \textbf{CodeBERT}       & 0.781          & 0.800          & 0.564           & 0.641          & 0.659          & 0.712          & 0.772          & 0.793                               \\ 
        \cline{2-10}
                                                 & \textbf{GraphCodeBERT}  & 0.784          & 0.804          & 0.537           & 0.652          & 0.663          & 0.693          & 0.803          & 0.829                               \\ 
        \cline{2-10}
                                                 & \textbf{seBERT}         & 0.754          & 0.777          & 0.525           & 0.624          & 0.635          & 0.678          & 0.749          & 0.772                               \\ 
        \cline{2-10}
                                                 & \textbf{CodeT5}         & 0.779          & 0.796          & 0.544           & 0.643          & 0.651          & 0.693          & 0.786          & 0.809          \\ 
        \cline{2-10}
                                                 & \textbf{PLBart}         & 0.762 & 0.782& 0.521  & 0.619 & 0.633 & 0.679 & 0.768 & 0.795  \\ 
        \cline{2-10}
                                                 & \textbf{CodeGen}        & 0.721               &     0.735           &  0.556               & 0.627               & 0.636               & 0.660               & 0.705               & 0.718             \\ 
        \hline
        \textbf{Our Model}                       & \textbf{SOBERT}         & \textbf{0.807(+2.9\%)} & \textbf{0.826(+2.7\%)} & \textbf{0.579}  & \textbf{0.678} & \textbf{0.684} & \textbf{0.732} & \textbf{0.801} & \textbf{0.832}                      \\
        \hline
        \end{tabular}
        }
        \end{table*}

\begin{table*}
    \centering
    \caption{Experiment Result for Relatedness Prediction Task}
    \label{tab:relate}
    \footnotesize
    \begin{tabular}{c|c|c|c|c} 
    \hline
    \textbf{Group}                           & \textbf{Representation} & \textbf{F1-Score} & \textbf{Precision} & \textbf{Recall}  \\ 
    \hline
    \textbf{SOTA}                            & \textbf{ASIM}           & 0.785             & 0.785              & 0.785            \\ 
    \hline
    \multirow{2}{*}{\textbf{SO-Specific}}    & \textbf{Post2Vec}       & 0.768             & 0.768              & 0.768            \\ 
    \cline{2-5}
                                             & \textbf{BERTOverflow}   & 0.697             & 0.697              & 0.697            \\ 
    \hline
    \multirow{3}{*}{\textbf{General Domain}} & \textbf{RoBERTa}        & 0.787             & 0.787              & 0.787            \\ 
    \cline{2-5}
                                             & \textbf{Longformer}     & 0.786             & 0.786              & 0.786            \\ 
    \cline{2-5}
                                             & \textbf{GPT2}           & 0.765             & 0.765              & 0.765            \\ 
    \hline
    \multirow{6}{*}{\textbf{SE domain}}      & \textbf{CodeBERT}       & 0.803             & 0.803              & 0.803            \\ 
    \cline{2-5}
                                             & \textbf{GraphCodeBERT}  & 0.801             & 0.801              & 0.801            \\ 
    \cline{2-5}
                                             & \textbf{seBERT}         & 0.799             & 0.799              & 0.799            \\ 
    \cline{2-5}
                                             & \textbf{CodeT5}         & 0.784             & 0.784              & 0.784            \\ 
    \cline{2-5}
                                             & \textbf{PLBart}         & 0.770             & 0.770              & 0.770            \\ 
    \cline{2-5}
                                             & \textbf{CodeGen}        & 0.765             & 0.765              & 0.765            \\ 
    \hline
    \textbf{Our Model}                       & \textbf{SOBERT}     & \textbf{0.825(+2.7\%)}            & \textbf{0.825(+2.7\%)}              & \textbf{0.825(+2.7\%)}            \\
    \hline
    \end{tabular}
    \end{table*}

\subsection*{RQ1: How effective are the existing \so post representation models?}
The experimental results of our tag recommendation experiments are summarized in Table \ref{tab:tag}. PTM4Tag achieves a performance of 0.417, 0.805, and 0.526 in terms of Precision@5, Recall@5, and F1-score@5, respectively.
However, the extra inclusion of Post2Vec lowers the scores to 0.416, 0.804, and 0.525, respectively. In contrast, BERTOverflow struggles in the task with surprisingly low scores of 0.083, 0.163, and 0.105.

For API recommendation, as shown in Table \ref{tab:api}, combining Post2Vec with the state-of-the-art approach CLEAR also fails to boost the performance. CLEAR itself obtains an MRR score of 0.739 and MAP score of 0.753. Yet, with the integration of Post2Vec, these values diminish slightly to 0.735 and 0.745, respectively. Notably, BERTOverflow achieves scores of 0.753 in MRR and 0.778 in MAP.

In the relatedness prediction task, as detailed in Table \ref{tab:relate}, integrating Post2Vec with ASIM leads to a minor decrease in the F1-score, moving from 0.785 to 0.768. BERTOverflow, on the other hand, lags behind ASIM with an F1-score of 0.697.

Overall, Post2Vec can not enhance the performance of the state-of-the-art solutions across the evaluated downstream tasks. Furthermore, BERTOverflow demonstrates poor results in classification tasks and only achieves comparable performance with the state-of-the-art solution in API recommendation.

\begin{tcolorbox}
    \textbf{Answer to RQ1}: The existing \so representation methods fail to improve state-of-the-art performance in the three evaluated downstream tasks.
\end{tcolorbox}

\subsection*{RQ2: How effective are the popular transformer-based language models for the targeted downstream tasks?}

In the tag recommendation task, as demonstrated in Table \ref{tab:tag}, the state-of-the-art approach PTM4Tag is outperformed by numerous transformer-based pre-trained representation models.  While the F1-score@5 of PTM4Tag is 0.526, PLBart achieves an F1-score@5 of 0.532, which makes it the best transformer-based pre-trained model. In contrast, seBERT significantly underperformed in this task, with an F1-score@5 of only 0.105. 

Table \ref{tab:api} shows that CLEAR is no longer the best-performing method in API recommendation. Replacing the embedding of Distilled RoBERTa in the original design of CLEAR with other transformer-based language models increases the performance. Particularly,
GraphCodeBERT boosts the performance of CLEAR by 3.8\% and 5.0\% in terms of MRR and MAP. For Precision@1,3,5 and Recall@1,3,5, GraphCodeBERT outperforms CLEAR by 6.7\% to 22.0\%. The worst representation model is CodeGen, which achieves MRR and MAP of 0.721 and 0.735. Meanwhile, both CodeBERT and GraphCodeBERT can surpass an MRR of 0.78.

The results of the relatedness prediction task are presented in Table \ref{tab:relate}. We observe that ASIM, the state-of-the-art technique in relatedness prediction, is outperformed by other transformer-based language models. While ASIM achieves a score of 0.785 in the F1-score, CodeBERT drives forward the state-of-the-art performance by 2.3\% with an F1-score of 0.803. RoBERTa, GraphCodeBERT, Longformer, and seBERT have an F1-score of 0.787, 0.801, 0.786, and 0.799, all outperforming ASIM.

Overall, models like CodeBERT, RoBERTa, and GraphCodeBERT can consistently give promising representations in all three tasks, proving their generalizability and effectiveness in a wide range of SE-related tasks.

\begin{tcolorbox}
    \textbf{Answer to RQ2}: Representations generated by CodeBERT,  RoBERTa, and GraphCodeBERT consistently outperform each state-of-the-art technique from the targeted downstream tasks.
    However, none of the models can always be the best performer. 
    
\end{tcolorbox}

\subsection*{RQ3: Is further pre-training on \so data helpful in building a better model?}

Our experimental results show that there is no ``one-size-fits-all" model in representing \so posts, which could consistently outperform others in the considered tasks. Such a phenomenon delivers an intuition that there is an improvement opportunity in the representation technique for \so.
Based on the common practice that a second phase of in-domain pre-training leads to performance gains~\cite{gururangan2020don}, we conduct additional pre-training for a transformer-based model (i.e., CodeBERT) with the \so dataset. We name it SOBERT.

\noindent\textbf{Pre-training Details:} We have leveraged the \so dump dated June 2023, which has 23 million question posts as the training corpus. 
The raw dataset has a size of approximately 70G. 
We also make sure that there is no overlapping between the test/validation datasets of three downstream tasks and the pre-training data of SOBERT.
Many previous works have removed the code snippets of a \so post during the pre-processing stage~\cite{zhou2019deep,tagdc}.   
According to the statistics conducted by Xu et al. ~\cite{post2vec}, more than 70\% of the \so contains at least one code snippet. As a result, the removal of code snippets would result in losing a significant of information, and they should be considered to learn an effective post representation. As the code snippets within the body of a post are enclosed in HTML tags \texttt{<pre><code>} and \texttt{</code></pre>}, we cleaned the redundant HTML tags with regular expression
\texttt{<pre><code>([\textbackslash{}s\textbackslash{}S]*?)<//code><//pre>}.
As a result, the pre-training data of SOBERT contains the title, body, and code snippets of a Stack Overflow question post.
We have initialized SOBERT based on the checkpoint of the CodeBERT model and pre-trained SOBERT using the MLM objective with a standard masking rate of 15\%. The batch size is set as 256, and the learning rate is 1e-4. The training process takes around 100 hours for eight Nvidia V100 GPUs with 16 GB of memory to complete. The detailed code used is included in the replication package provided. 

The experimental results show that SOBERT achieves the best performance for every downstream task. 
For tag recommendation, SOBERT achieves an F1-score@5 of 0.551 and beats PLBart by 3.4\%; for API recommendation, SOBERT performs with 0.807 in terms of MRR and outperforms GraphCodeBERT by 2.9\%.; and for relatedness prediction, it accomplishes an F1-score of 0.824 and outperforms CodeBERT by 2.7\%. 

We conduct the Wilcoxon Signed Rank~\cite{conover1999practical} at a 95\% significance level (i.e., p-value $<$ 0.05) and calculate Cliff's delta~\cite{cliff2014ordinal} on the paired data corresponding to SOBERT and the best-performing competing representation model in each task (i.e., PLBart in tag recommendation, CodeBERT in relatedness prediction, and GraphCodeBERT in API recommendation). The significance test has been conducted on the values of evaluation metrics (F1-score@5 in tag recommendation, F1-score in relatedness prediction, MRR in API recommendation).
For Cliff's delta, we consider delta that are less than 0.147, between 0.147 and 0.33, between 0.33 and 0.474, and
above 0.474 as Negligible (N), Small (S), Medium (M), and Large (L) effect size, respectively following previous literature~\cite{cliff2014ordinal}.
We observe that SOBERT significantly (p-value $<$ 0.05) and substantially (Cliff’s delta is 0.31–0.55) outperforms the comparing model. 

\begin{tcolorbox}
    \textbf{Answer to RQ3}: Further pre-training with the \so data yields better representation in modelling \so posts. SOBERT consistently achieves state-of-the-art performance in all the targeted downstream tasks.
\end{tcolorbox}

%% file: sections/discussion.tex
\section{discussion}
\label{sec:discussion}

\subsection{Lessons Learned}
\subsubsection*{\textbf{Lesson \#1}}
\textbf{\textit{Incorporating post embeddings from an external approach does not boost the performance of neural network models.}} 

Xu et al.~\cite{post2vec} demonstrated that appending the distributed post representation learned by Post2Vec to the manually crafted feature vector can increase the performance of traditional machine learning algorithms, for example, Support Vector Machine~\cite{xu2018prediction} and Random Forest~\cite{beyer2018automatically}, in a set of \so-related tasks. 
However, these benefits are not observed for the state-of-the-art techniques that are based on deep neural networks. 
This is potentially caused by the design of neural networks that automatically extract feature vectors and continuously optimize the representations. 
It indicates that deep neural networks may lose the effectiveness of external embeddings while optimizing the parameters of the feature extractor.

\subsubsection*{\textbf{Lesson \#2}}

\textbf{\textit{Models with broader background knowledge derive better results than those with specific knowledge.}}

Intuitively, BERTOverflow is expected to produce the desired \so post representation as it is specifically designed for \so data.
A major difference between BERTOverflow and others is the vocabulary. 
As BERTOverflow is pre-trained from scratch with the \so data, its vocabulary should be more suitable for modeling \so posts than general domain models.

Surprisingly, our experiment results show that other transformer-based language models outperform BERTOverflow by a substantial margin across all three tasks. It gives an extremely poor performance in the tag recommendation task. By inspecting the prediction results of BERTOverflow in the tag prediction task, we notice that the top-5 predictions made by BERTOverflow are always the most frequent tags (`python', `java', `c\#', `java-script', and `android') from the dataset. We observe seBERT has a similar performance as BERTOverflow in the tag recommendation task.

We hypothesize that the poor performance of BERTOverflow and seBERT is because these models lack a sufficient amount of pre-training to perform well. Since seBERT and BERTOverflow are trained from scratch, they require much more pre-training effort than continued pre-training with existing models.
To prove this concept, we perform additional pre-training on BERTOverflow with the same dataset as SOBERT.
The further pre-training was done with the same hyper-parameters as SOBERT, and it took 23 hours on four 16GB Nvidia V100 GPUs to complete.
We denote this new model as BERTOverflow$_{\text{NEW}}$.\footnote{Please note that we could not apply further pre-training to seBERT due to the constraints of limited resources to handle BERT$_{\text{Large}}$ architecture.}

To independently evaluate the effect of domain-specific vocabulary, we train additional two transformer-based models from scratch. Both models contain 12 layers of encoder modules, but one with the vocabulary of BERTOverflow and the other with the vocabulary of RoBERTa. 
The pre-training process is the same as for BERTOverflow$_{\text{NEW}}$. We refer to these two models as BERTOverflow$_{\text{vocab}}$, and RoBERT$_{\text{vocab}}$. 
From Table \ref{tab:bertoverflow}, we observe the significant performance improvements of  BERTOverflow$_{\text{NEW}}$ compared to BERTOverflow. We also observe that the performance of BERTOverflow$_{\text{vocab}}$ and RoBERTa$_{\text{vocab}}$ are very similar to each other. Overall, our experiments have shown that domain-specific vocabulary has a negligible effect on all three tasks. The more important factor tends to be the amount of pre-training.
Moreover, pre-training from scratch is commonly considered an expensive process. Initializing new representation models based on the checkpoint of a popular model reduces the risk, and the tag recommendation task is a good indicator to demonstrate the generalizability and the sufficiency of pre-training of transformer-based representation models.

For a more comprehensive insight, we analyze the lengths of tokenized text for different representation models on the datasets used in this paper.  Table \ref{tab:tag-tokenize}, \ref{tab:api-tokenize}, and \ref{tab:relate-tokenize} show the statistics for the lengths of different tokenizers. 
Given that representation models like CodeBERT, GraphCodeBERT, Longformer, GPT2, and CodeGen share the same tokenizer as RoBERTa, we omit these representation models from this analysis. 
We notice that PLBart's tokenizer can generate the shortest tokenizations across all datasets. Interestingly, even though BERTOverflow and seBERT are being developed with SE-specific vocabularies, 
the lengths of their tokenized text are almost the same as RoBERTa's. For example, BERTOverflow's average tokenization length in the tag recommendation task is 250, while RoBERTa's is slightly higher at 254.7. 

\begin{table}[t]
  \caption{Comparison of the results from different tokenizers on the dataset of tag recommendation.}
  \label{tab:tag-tokenize}
    \centering

  \begin{tabular}{c|c|c|c|c|c|c|c}
  \hline
  \textbf{Model} & \textbf{Mean} & \textbf{Min} & \textbf{25\%} & \textbf{50\%} & \textbf{75\%} & \textbf{Max} & \textbf{Longer than 512} \\
  \hline
  \textbf{CodeT5} & 253.3 & 8 & 93 & 158 & 278 & 27449 & 9.3\% \\
  \textbf{RoBERTa} & 254.7 & 9 & 92 & 157 & 279 & 27437 & 9.4\% \\
  \textbf{PLBart} & 239 & 8 & 90 & 152 & 266 & 19467 & 8.4\% \\
  \textbf{BERTOverflow} & 250 & 8 & 90 & 156 & 278 & 27596 & 9.3\% \\
  \textbf{seBERT} & 255.4 & 8 & 92 & 158 & 283 & 27600 & 9.7\% \\
  \hline
  \end{tabular}
\end{table}

\begin{table}[]
  \centering
  \caption{Comparison of the results from different tokenizers on the dataset of API recommendation.}
  \label{tab:api-tokenize} 
  \begin{tabular}{c|c|c|c|c|c|c}
    \hline
    \textbf{Model} & \textbf{Mean} & \textbf{Min} & \textbf{25\%} & \textbf{50\%} & \textbf{75\%} & \textbf{Max} \\
    \hline
    \textbf{CodeT5} & 11.0 & 4 & 8 & 10 & 14 & 53 \\
    \textbf{RoBERTa} & 11.3 & 4 & 8 & 11 & 14 & 49 \\
    \textbf{PLBart} & 10.5 & 3 & 7 & 10 & 13 & 51 \\
    \textbf{BERTOverflow} & 10.3 & 4 & 7 & 10 & 13 & 48 \\
    \textbf{seBERT} & 10.6 & 4 & 7 & 10 & 13 & 48 \\
    \hline
  \end{tabular}
\end{table}

\begin{table}[]
  \centering
  \caption{Comparison of the results from different tokenizers on the dataset of relatedness prediction.}
  \label{tab:relate-tokenize}
  \begin{tabular}{c|c|c|c|c|c|c|c}
    \hline
  \textbf{Model Name}   & \textbf{Mean} & \textbf{Min} & \textbf{25\%} & \textbf{50\%} & \textbf{75\%} & \textbf{Max} & \textbf{Longer than 512} \\
  \hline
  \textbf{CodeT5}       & 1826.6        & 102          & 919           & 1407          & 2213          & 33989        & 94.6\%                   \\
  \textbf{RoBERTa}      & 1897.7        & 99           & 944           & 1454          & 2301          & 34540        & 94.8\%                   \\
  \textbf{PLBart}       & 1717.9        & 96           & 877           & 1340          & 2090          & 32995        & 93.7\%                   \\
  \textbf{BERTOverflow} & 1862.7        & 98           & 937           & 1444          & 2264          & 41346        & 94.7\%                   \\
  \textbf{seBERT}       & 1917.4        & 98           & 961           & 1485          & 2331          & 41750        & 95.0\%    \\
  \hline              
  \end{tabular}
  \end{table}

\begin{table*}[]
\centering
\caption{Results for variants of BERTOverflow, RoBERTa, and Longformer}

\label{tab:bertoverflow}
\resizebox{\textwidth}{!}{
\begin{tabular}{c|ccc|cc|ccc}
\hline
                          & \multicolumn{3}{c|}{\textbf{Tag Recommendation}}                                      & \multicolumn{2}{c|}{\textbf{API Recommendation}} & \multicolumn{3}{c}{\textbf{Relatedness Prediction}}                             \\ \hline
                          & \multicolumn{1}{c|}{\textbf{P@5}} & \multicolumn{1}{c|}{\textbf{R@5}} & \textbf{F1@5} & \multicolumn{1}{c|}{\textbf{MRR}} & \textbf{MAP} & \multicolumn{1}{c|}{\textbf{P}} & \multicolumn{1}{c|}{\textbf{R}} & \textbf{F1} \\ \hline
\textbf{BERTOverflow}     & \multicolumn{1}{c|}{0.083}        & \multicolumn{1}{c|}{0.163}        & 0.105         & \multicolumn{1}{c|}{0.753}        & 0.778        & \multicolumn{1}{c|}{0.697}      & \multicolumn{1}{c|}{0.697}      & 0.697       \\ \hline
\textbf{BERTOverflow-New} & \multicolumn{1}{c|}{0.411}        & \multicolumn{1}{c|}{0.791}        & 0.519         & \multicolumn{1}{c|}{0.779}        & 0.793        & \multicolumn{1}{c|}{0.789}      & \multicolumn{1}{c|}{0.789}      & 0.789       \\ \hline
\textbf{BERTOverflow-vocab} & \multicolumn{1}{c|}{0.411}        & \multicolumn{1}{c|}{0.790}        & 0.518         & \multicolumn{1}{c|}{0.771}        & 0.785        & \multicolumn{1}{c|}{0.788}      & \multicolumn{1}{c|}{0.788}      & 0.788       \\ \hline
\textbf{RoBERTa-vocab} & \multicolumn{1}{c|}{0.412}        & \multicolumn{1}{c|}{0.794}        & 0.520         & \multicolumn{1}{c|}{0.778}        & 0.792        & \multicolumn{1}{c|}{0.793}      & \multicolumn{1}{c|}{0.793}      & 0.793       \\ \hline

\textbf{Longformer-512}   & \multicolumn{1}{c|}{0.397}        & \multicolumn{1}{c|}{0.768}        & 0.502         & \multicolumn{1}{c|}{0.768}        & 0.783        & \multicolumn{1}{c|}{0.785}      & \multicolumn{1}{c|}{0.785}      & 0.785       \\ \hline
\textbf{Longformer-1024}  & \multicolumn{1}{c|}{0.397}        & \multicolumn{1}{c|}{0.769}        & 0.502         & \multicolumn{1}{c|}{0.767}        & 0.782        & \multicolumn{1}{c|}{0.786}      & \multicolumn{1}{c|}{0.786}      & 0.786       \\ \hline
\end{tabular}%
}
\end{table*}

\subsubsection*{\textbf{Lesson \#3}}
\textbf{\textit{Despite considering a longer input length, Longformer does not produce better representations for posts.}}

Conventional transformer-based models like CodeBERT and RoBERTa cannot handle long sequences due to the quadratic complexity of the self-attention mechanism~\cite{transformer} and accept a maximum of 512 tokens as the input. From Table \ref{tab:tag-tokenize}, \ref{tab:api-tokenize}, and \ref{tab:relate-tokenize}, we can observe that the ratios of data that are longer than 512 tokens are approximately 9\%, 0\%, and 94\% in tag recommendation, API recommendation, and relatedness prediction, respectively.

In Table \ref{tab:api-tokenize}, we can see that the dataset of API recommendation has a short length, where the longest tokenized text has a length of 53. This is because only the title of a post is considered in this task. As Longformer is implemented with a simplified attention mechanism (introduced in Section 2), which only gives its advantage in handling long text, this explains why CodeBERT, and RoBERTa outperform Longformer in API recommendation. 

From Table \ref{tab:tag-tokenize} and \ref{tab:relate-tokenize}, we can see that both dataset contains data samples that are longer than the 512 limit. Especially in related prediction (Table \ref{tab:relate-tokenize}), the average length of each knowledge unit (KU) is more than 1800 tokens. 
The lengthy text is because each KU consists of question posts and a set of corresponding answer posts. Surprisingly, Longformer fails to perform better than the other model that belongs to the general domain (i.e., RoBERTa) as well as models from the SE domain, even though it takes much longer input in this task.

We further compare the performance of Longformer by varying the input size considering the first 512 and 1,024 tokens.
The additional experimental results are shown in Table~\ref{tab:bertoverflow}. 
These additional settings do not differ in performance. 
It indicates that diversifying the input size does not affect Longformer's performance on post representation.
A potential interpretation would be the important features for representing \so posts lie in the first part of each post (e.g., Title serves as a succinct summary of the post).
It is not worth trying Longformer unless one strictly needs the entire content of \so posts.

\begin{table*}[]
\centering
\caption{Examples of predictions made by CodeBERT and SOBERT in the tag recommendation task}

\label{tab:tag-sample}
\resizebox{\textwidth}{!}{%
\begin{tabular}{c|cccc}
\toprule
\textbf{\begin{tabular}[c]{@{}c@{}}Post\\ ID\end{tabular}} &
  \textbf{\begin{tabular}[c]{@{}c@{}}Post\\ Title\end{tabular}} &
  \textbf{\begin{tabular}[c]{@{}c@{}}CodeBERT\\ Tag Prediction\end{tabular}} &
  \textbf{\begin{tabular}[c]{@{}c@{}}SOBERT\\ Tag Prediction\end{tabular}} &
  \textbf{True Tag} \\ 
\midrule
\textbf{13202867} &
  Fixed size of Flexslider &
  \begin{tabular}[c]{@{}c@{}}apache-flex, frameworks, \\ ios, swift, xcode\end{tabular} &
  \begin{tabular}[c]{@{}c@{}}css, html, image,  \\ javascript, jquery\end{tabular} &
  css, html, javascript \\ 
\textbf{30434343} &
  \begin{tabular}[c]{@{}c@{}}What is the right way to \\ typecheck dependent \\ lambda abstraction \\ using `bound'?\end{tabular} &
  \begin{tabular}[c]{@{}c@{}}.net, binding, c\#, \\ lambda, type-inference\end{tabular} &
  \begin{tabular}[c]{@{}c@{}}functional-programming, haskell, \\ lambda, type-inference, types\end{tabular} &
  haskell \\ 
\textbf{17849870} &
  Closed type classes &
  \begin{tabular}[c]{@{}c@{}}.net, c++, d, \\ f\#, performance\end{tabular} &
  \begin{tabular}[c]{@{}c@{}}applicative, ghc, haskell, \\ typeclass, types\end{tabular} &
  \begin{tabular}[c]{@{}c@{}}haskell, static-analysis, \\ typeclass, types\end{tabular} \\ 
\bottomrule  
\end{tabular}%
}
\end{table*}

\subsubsection*{\textbf{Lesson \#4}}

\textbf{\textit{We advocate future studies related to \so consider the SOBERT as the underlying baseline.}} 

Our experiment results demonstrate that further pre-training based on in-domain data leads to better \so post representation. By initializing SOBERT with the CodeBERT checkpoint and performing further pre-training on \so data, we have noticed that SOBERT consistently outperforms the original CodeBERT and produces new state-of-the-art performance for all three tasks.

In Table \ref{tab:tag-sample}, we present three examples of the prediction results of CodeBERT and SOBERT for the tag recommendation task. We observe that CodeBERT is making wrong predictions like ``.net'' and ``c\#'' when the question is about ``haskell'', while SOBERT is capable of making the correct predictions. CodeBERT may lack knowledge of programming languages like Haskell and Lua since it is pre-trained on artifacts from Python, Java, JavaScript, PHP, Ruby, and Go. Taking the \so post with ID 13202867 as another example, the question is about Flexslider, a jQuery slider plugin. In the given example, SOBERT could successfully make connections to tags like `jQuery' and `css' while CodeBERT struggles to give meaningful predictions. 

Overall, by continuing the pre-training process on \so data, SOBERT outperforms CodeBERT in three popular \so-related tasks. 
We advocate future studies to consider SOBERT as their underlying baseline. To facilitate the usage of the SOBERT proposed in this work, we plan to release it to HuggingFace\footnote{\url{https://huggingface.co/}} so that it can be used by simply calling the interface.

\subsection{Threats to Validity}

\newcommand{\hochkomma}{$^{,}$}

\noindent\textbf{Threats to internal validity.} To ensure the correct implementation of the baseline methods (i.e., Post2Vec, PTM4Tag, CLEAR, and ASIM), we reused the replication package released by the original authors.\footnote{\url{https://github.com/maxxbw54/Post2Vec}}\hochkomma\footnote{\url{https://github.com/soarsmu/PTM4Tag}}\hochkomma\footnote{\url{https://github.com/Moshiii/CLEAR-replication}}\hochkomma\footnote{\url{https://github.com/Anonymousmsr/ASIM}} 
When investigating the effectiveness of various pre-trained models, we used the implementation of each from the popular open-source community \textit{HuggingFace}.
Another threat to the internal validity is the hyperparameter setting we used to pre-train SOBERT and fine-tune the representation models. To mitigate this threat, the leveraged hyper-parameters in this paper 
in both the pre-training and fine-tuning phases were reported in other prior reputable literature as recommended or optimal~\cite{codebert, Wei2022CLEARCL, roberta}. 

\noindent\textbf{Threats to external validity.}
One threat to external validity relates our results may not generalize to those newly emerging topics or other \so-related downstream tasks. We have minimized this threat by considering multiple downstream tasks. 

\noindent\textbf{Threats to construct validity.} We reuse the same evaluation metrics in our baseline methods~\cite{ptm4tag, Wei2022CLEARCL, Pei2021AttentionbasedMF}. To further reduce the risk, we conduct the Wilcoxon signed-rank statistical hypothesis test and Cliff's delta to check whether the output between the two competing approaches is significant and substential.

%% file: sections/relate.tex
\section{related work}
\label{sec:relate}
In this section, we review two lines of research that most relate to our work: pre-trained models for SE and mining Stack Overflow posts.

\subsection{Pre-trained Models for Software Engineering}
Inspired by the success of transformer-based pre-trained models in NLP, there is an increasing research interest in exploring pre-training tasks and applying pre-trained models for SE~\cite{contracode,lu2021codexglue,codeT5,zhang2020sentiment,tracebert, CiniselliCPMAPP22, MastropaoloSCNP21, TufanoMMPPB22, MastropaoloPB22}. One set of research focuses on learning semantic and contextual representations of source code; after pre-training, these models can be fine-tuned to solve SE downstream tasks.
ContraCode~\cite{contracode} is an encoder-only model that uses a contrastive pre-training task to learn code functionality. It classifies JavaScript programs into positive pairs (i.e., functionally similar) and negative pairs (i.e., functionally dissimilar). CCBERT~\cite{ccbert} a transformer-based encoder-only PTM that learns a generic representation of code changes. 
CodeGPT~\cite{lu2021codexglue} is a pure decoder model trained in programming languages. It leverages the Python and Java corpora from the CodeSearchNet dataset~\cite{codesearchnet}.

Another set of research focuses on leveraging the transformer-based model to automate SE challenges~\cite{zhang2020sentiment,tracebert,CiniselliCPMAPP22,xinICPC}.
Lin et al.~\cite{tracebert} find that BERT can boost the performance of traceability tasks in open-source projects.
They investigate three BERT architectures, i.e., Single-BERT, Siamese-BERT, and Twin-BERT.
The results indicate that the single-BERT can generate the most accurate links, while a Siamese-BERT architecture produced comparable effectiveness with significantly better efficiency.
Ciniselli et al.~\cite{CiniselliCPMAPP22} investigate the potential of transformer-based models in the code completion task, ranging from single token prediction to the prediction of the entire code blocks.
Experiments are conducted on several variants of two popular transformer-based models, namely RoBERTa and T5. The experimental results demonstrate that T5 is the most effective in supporting code completion. 
T5 variants can achieve prediction accuracies up to 29\% for whole block prediction and 69\% for token-level prediction. 
Mastropaolo et al.~\cite{MastropaoloSCNP21} further explore the effusiveness of T5 for bug fixing, injecting code mutants, generating assert statements, and code summarization. The study finds that the T5 can outperform the previous state-of-the-art deep learning-based approaches for those tasks.
Tufano et al.~\cite{TufanoMMPPB22} perform an empirical evaluation of the T5 model in automating the code review process. The experiments are performed on a much
larger and realistic code review dataset, and the T5-based model outperforms previous deep learning models in this task.
Mastropaolo et al.~\cite{MastropaoloPB22} present LANCE, a log statement recommendation system using the transformer architecture. LANCE can accurately determine where to place a log statement at 65.9\% of the time, select the correct log level at 66.2\% of the time, and generate a fully accurate log statement with a relevant message in 15.2\% of instances.

Different from these works, we focus on a comprehensive set of \so-related tasks in this paper. 
In addition to fine-tuning the transformer-based pre-trained representation models, we also further pre-trained SOBERT on Stack Overflow data.

\subsection{Mining Stack Overflow Posts}
We address tag recommendation~\cite{ptm4tag, post2vec}, API recommendation~\cite{Wei2022CLEARCL, biker}, and relatedness prediction~\cite{xu2018prediction,Pei2021AttentionbasedMF} in this work.
Others also explored other tasks for mining Stack Overflow posts to support software developers, such as post recommendation~\cite{postfinder}, multi-answer summarization~\cite{answerbot}, and controversial discussions~\cite{ren2019discovering}.

Rubei et al.~\cite{postfinder} propose an approach named PostFinder, which aims to retrieve Stack Overflow posts that are relevant to API function calls that have been invoked.
They make use of Apache Lucene to index the textual content and code in Stack Overflow to improve efficiency.
In both the data collection and query phase, they make use of the data available at hand to optimize the search process.
Specifically, they retrieve and augment posts with additional data to make them more exposed to queries.
Besides, they boost the context code to construct a query that contains the essential information to match the stored indexes.

Xu et al.~\cite{answerbot} investigate the multi-answer posts summarization task for a given input question, which aims to help developers get the key points of several answer posts before they dive into the details of the results.
They propose an approach \textit{AnswerBot}, which contains three main steps, i.e., relevant question retrieval, useful answer paragraph selection, and diverse answer summary generation.

Ren et al.~\cite{ren2019discovering} investigate the controversial discussions in Stack Overflow.
They find that there is a large scale of controversies in Stack Overflow, which indicates that many answers are wrong, less optimal, and out-of-date.
Our work and their work are complementary to each other, and all aim to boost automation in understanding and utilizing Stack Overflow contents.

%% file: sections/conclusion.tex
\section{Conclusion and Future Work}
\label{sec:conclusion}

In this paper, we empirically study the effectiveness of varying techniques for modeling \so posts, including approaches that are specially designed for \so posts (i.e., Post2Vec and BERTOverflow),
SE domain representation models (i.e., CodeBERT, GraphCodeBERT, seBERT, CodeT5, PLBart, CodeGen) and general domain representation models (i.e., RoBERTa, LongFormer, and GPT2). We evaluate the performance of these representation models on three popular and representative \so-related tasks, which are tag recommendation, API recommendation, and relatedness prediction.

Our experimental results show that Post2Vec is unable to enhance the representations that are automatically extracted by deep learning-based methods and BERTOverflow performs surprisingly worse than other transformer-based language models. Furthermore, there does not exist one representation technique that could consistently outperform other representation models. Our findings indicate the current research gap in representing \so posts. Thus, we propose SOBERT with a simple-yet-effective strategy. We pre-train SOBERT with the posts from \so. As a result, SOBERT improves the performance of the original CodeBERT and consistently outperforms other models on all three tasks, confirming that further pre-training on \so data helps build \so representation. 

In the future, we would also extend our research to other SQA sites, such as AskUbuntu\footnote{\url{https://askubuntu.com/}}. Moreover, we would also consider other \so-related downstream tasks into account in the future. 

\section{Data Availability}
The replication package of the data and code used in this paper is available at \textbf{\url{https://figshare.com/s/7f80db836305607b89f3}}.